\documentclass[twocolumn,showpacs,amsmath,amssymb]{revtex4}

\usepackage{graphicx} 
\usepackage{dcolumn}

\begin{document}

\title{A novel manifestation of $\alpha$-clustering:  
new '$\alpha$ + $^{208}$Pb' states in $^{212}$Po revealed by their 
enhanced E1 decays}

\author{A. Astier$^1$}
\author{P. Petkov$^{1,2}$}
\author{M.-G. Porquet$^1$}
\author{D.S. Delion$^{3,4}$}
\author{P. Schuck$^5$}

\affiliation{$^1$CSNSM, IN2P3-CNRS and Universit\'e Paris-Sud 
91405 Orsay, France\\
$^2$INRNE, BAS, 1784 Sofia, Bulgaria\\
$^3$Horia Hulubei National Institute of Physics and Nuclear 
Engineering  407 Atomistilor, 077125 Bucharest, Romania\\
$^4$ Academy of Romanian Scientists, 54 Splaiul Independentei  
050094 Bucharest, Romania\\
$^5$ IPN, IN2P3-CNRS and Universit\'e Paris-Sud  91406 Orsay, France\\
}

\date{\today}
\begin{abstract}
Excited states in $^{212}$Po were populated by $\alpha$ 
transfer using the $^{208}$Pb($^{18}$O, $^{14}$C) reaction and their
deexcitation $\gamma$-rays were studied with the Euroball array. 
Several 
levels were found to decay by a unique E1 transition 
(E$_\gamma <$ 1~MeV) populating the yrast state with the same spin 
value. Their lifetimes were 
measured by the DSAM method. The values, found in the range 
[0.1-1.4]~ps, lead to very enhanced transitions, 
B(E1) = 2$\times$10$^{-2}$~-~1$\times$10$^{-3}$ W.u.. These results 
are discussed in terms of an $\alpha$-cluster structure which gives
rise to states with non-natural parity values, provided that the 
composite system cannot rotate collectively, as expected in the 
'$\alpha$+$^{208}$Pb' case. Such states due to the oscillatory motion of
the $\alpha$-core distance are observed for the first time.   
\end{abstract}


\maketitle

At the early days in the development of nuclear theory, the $\alpha$ 
particle was considered as the basic building block of any nucleus, 
providing a simple explanation for the emission of $\alpha$'s by heavy 
nuclei or the fact that all the light nuclei with $A=4n$ have
higher binding energies per particle than any of their neighbours.
But many arguments~\cite{be36} were rapidly developed against this 
picture, which was almost completely abandoned to the benefit 
of single-particle description of nuclei based on the hypothesis of 
a common mean field for all nucleons. 

A strong revival of the 
$\alpha$-cluster model occurred in the 1960s when both experimental 
and theoretical studies revealed that the concept of 
$\alpha$-clustering is essential for the understanding of the structure 
of light nuclei. The states based on $\alpha$-particles (and other 
bound sub-structures) are not so much found in the ground states 
but rather observed as excited states close to the decay thresholds 
into clusters, as suggested by Ikeda~\cite{ik68}. In particular, 
the Hoyle state, i.e. the $0_2^+$-state at 7.65 MeV in $^{12}$C 
which has recently been interpreted as an $\alpha$-particle 
condensate~\cite{to01}, and other similar states in 
heavier $n\alpha$ nuclei, have attracted much renewed attention, 
see e.g.~\cite{sc07}. Also so-called nuclear 
molecules, such as the $\alpha$-core system rotating about 
its center of mass, have lately been 
studied intensely and
with great success (for a review, see~\cite{oe06}).

The persistence of $\alpha$-clustering in heavy nuclei is much less 
documented. This is even the case of $^{212}$Po, a typical nucleus 
with two protons and two neutrons outside the doubly-magic core, 
$^{208}$Pb. Up to now, there was no clear-cut evidence of its cluster
structure. Shell model (SM) configurations involving a few orbits account
reasonnably well for the excitation energies of its low-lying yrast 
states~\cite{po87}.
Nevertheless, the $\alpha$ width of its ground state is predicted more 
than one order of magnitude smaller than the experimental value. 
This has called
for a hybrid model comprising both shell and cluster 
configurations~\cite{va92}. Then the $\alpha$-decay energy and the
half-life are well reproduced and a large amount of $\alpha$-clustering is
found (30\%). Such a result suggests that the low-lying yrast 
spectrum of 
$^{212}$Po could also be explained in terms of an $\alpha$-$^{208}$Pb 
cluster model, that was explored in some theoretical works 
(see e.g.~\cite{oh95,bu96,de00}).

This Letter reports on the evidence of $\alpha$-clustering in 
$^{212}$Po  by means of several very enhanced E1 transitions which link 
excited states with non-natural parity to the yrast states having
the same spin values.  While such results are not expected from 
SM configurations, they can be explained in terms of 
'$\alpha$+$^{208}$Pb' structure. This represents a novel manifestation of 
$\alpha$-clustering, very different from that observed in 
light nuclei where the $\alpha$-core system can rotate collectively 
about its center of mass.

Excited states in $^{212}$Po were populated by $\alpha$
transfer using the 
$^{208}$Pb($^{18}$O, $^{14}$C) reaction. The $^{18}$O beam of 
85~MeV energy was provided by the Vivitron tandem of IReS 
(Strasbourg). The target of 100~mg/cm$^2$ $^{208}$Pb was thick 
enough to stop the recoiling nuclei as well as the $^{18}$O beam. 
The $\gamma$-rays were detected by the 71 Ge detectors of the 
Euroball IV 
array~\cite{si97}, i.e.
15 cluster detectors placed in the 
backward hemisphere with respect to the beam,
26 clover detectors located around 90$^\circ$, 
and 30 tapered single-crystal detectors located at 
forward angles. The 239 Ge crystals of the Euroball array 
could be grouped into 13 rings, 3 forward, 4 close to
90$^\circ$ and 6 backward, or into 2
groups at 39.3$^\circ$ and 76.6$^\circ$. 
Events were recorded on tape when at least 3
detectors fired in prompt coincidence, this led to 
a set of $\sim 4\times 10^9$ three- and higher-fold events. 
Various procedures have 
been used for the offline analysis in order to fully 
characterize the excited levels of $^{212}$Po  
(excitation energy, 
spin and parity, decay modes, and lifetime).

\begin{figure}[h!]
\begin{center}
\includegraphics[scale=0.37]{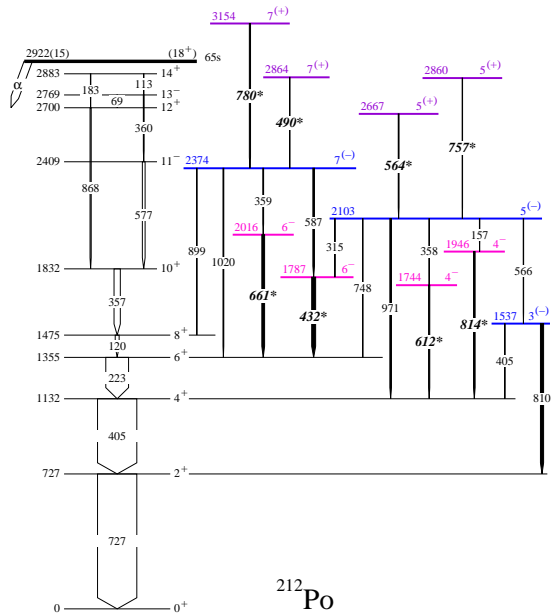}
\caption{
(Color online) Part of the level scheme of $^{212}_{~84}$Po$_{128}$ 
obtained in this
work. The long-lived isomeric state at 2922(15) keV excitation
energy, a pure $\alpha$-emitter not observed in the present work, is
drawn for the sake of completeness. The width of the
arrows is representative of the intensity of the
$\gamma$-rays. The transitions
marked with an asterisk are emitted by states with 
$\tau$ $<$ 1.4~ps (see Table 1). The colored states are also 
displayed in the bottom part of Fig.~\ref{etats_cluster}.
\label{schema_part_212Po}
}
\end{center}
\end{figure}

Both multi-gated spectra and three-dimensional
'cubes' have been built and analyzed with the 
Radware package \cite{ra95} in order to establish the level scheme. 
By gating on the known transitions~\cite{po87,po03} we have 
assigned about 50 new $\gamma$-rays to $^{212}$Po, de-exciting 
35 new excited states. About ten of them are located above
2.92~MeV, the energy of the (18$^+$) long-lived state. A partial 
level scheme showing the $\gamma$-decay of the yrast states 
and some of the levels which are the
object of this Letter is displayed in Fig.~\ref{schema_part_212Po}.
The complete level scheme will be published and discussed 
elsewhere~\cite{as09}. 
 
\begin{figure}[h!]
\begin{center}
\includegraphics[scale=0.15]{fit780.eps}
\includegraphics[scale=0.27]{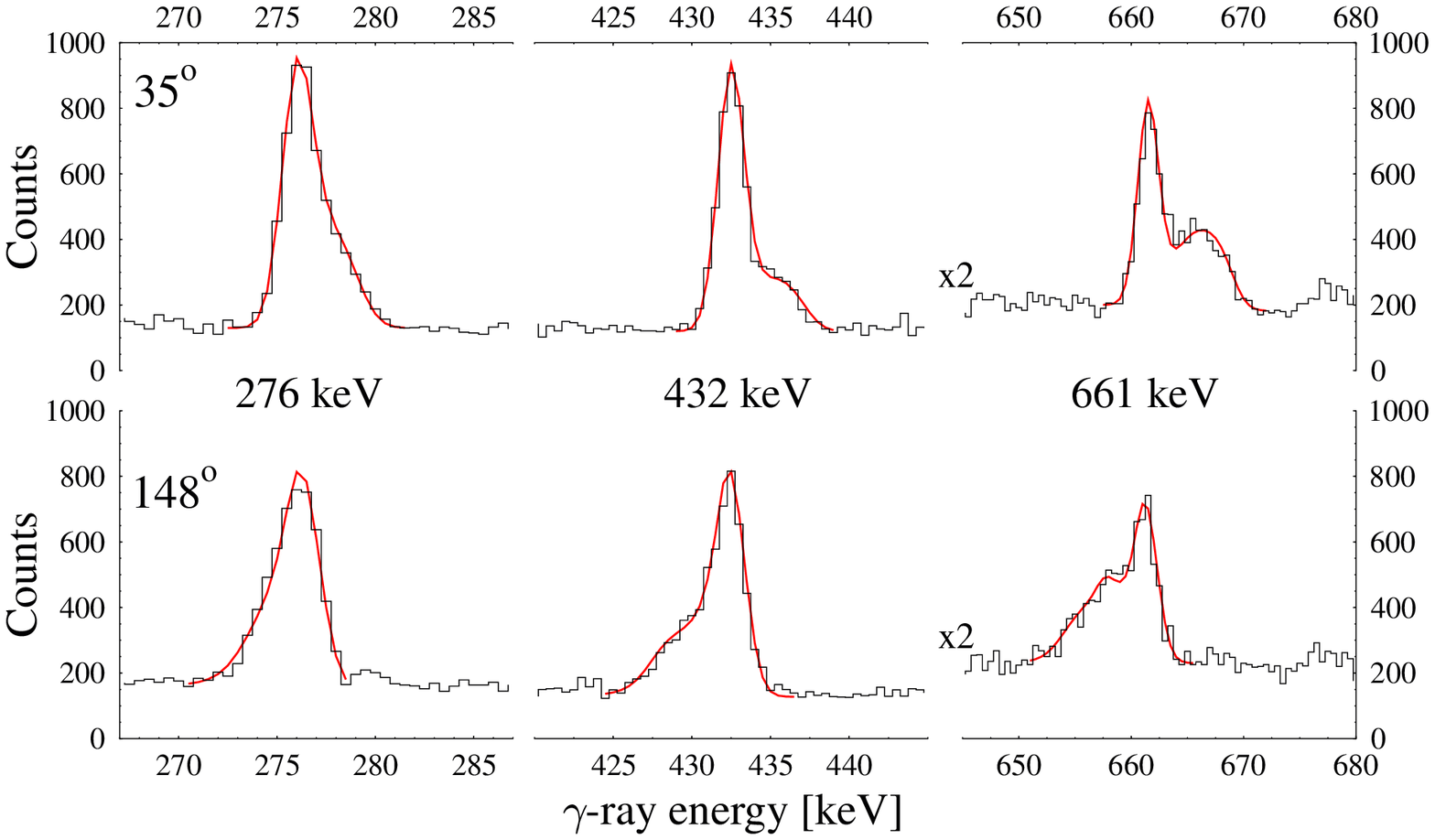}
\caption{(Color online) Top: Energy of the Doppler-shifted line around 780~keV as a 
function of the detection angle. The curve is the best fit 
obtained for a relative velocity $\frac{v}{c}$=1\% and a transition
energy of 780.4~keV. Bottom: Examples of line-shape analysis
at forward (35$^\circ$) and backward (148$^\circ$) angles. 
\label{doppler}  
}
\end{center}
\end{figure}

The stopping time of $^{212}$Po in the lead target is about one 
picosecond, thus it would have been expected that all the transitions lying 
in the low-energy part of the level scheme are emitted at rest.
Nevertheless we have found several $\gamma$-rays with E$_\gamma$ $<$
1~MeV which exhibit shifts and broadenings in energy due to the 
Doppler effect (some of them are marked by an asterisk in 
Fig.~\ref{schema_part_212Po}), meaning that they are 
emitted during the slowing down  and thus the corresponding excited states 
do have 
lifetimes $\lesssim$ 1~ps. 
The 780~keV transition, located in the high-energy part of the level
scheme, displays only shifted components. 
The value of its energy, measured as a function of the detector 
angle, is symmetric
around 90$^\circ$ (see the top part of Fig.~\ref{doppler}), 
showing that the $^{212}$Po nuclei recoil along the
beam axis, the best fit giving $\frac{v}{c}$=1\%.
The modulus of the  $^{212}$Po velocity proves 
that the  $^{14}$C ejectiles are also emitted along the beam axis,
but in the {\it backward} direction (in perfect agreement with
previous results~\cite{vi77,bo81}). Thus the $\alpha$
particle is transferred almost at rest.  
 
The lifetimes of seven excited levels were 
determined using the Doppler-shift 
attenuation method (DSAM), which is based on the time-correlation 
between the slowing down of the recoiling ion
and the decay of the nuclear level of interest (cf. e.g. 
Ref.~\cite{al78}). The data analysis was performed using a
standard procedure (cf. e.g. Ref.~\cite{Orce04}).
For the description of the slowing-down process via Monte-Carlo 
methods
we used a modified version of the program 
DESASTOP~\cite{pe98,wi83}.
The line-shapes of the  transitions to be analyzed were obtained 
using the coincidence matrices by setting
gates on fully stopped $\gamma$-ray peaks, belonging to transitions 
which depopulate levels lying below the level of interest. 
Examples of line-shape analysis
are displayed in the bottom part of Fig.~\ref{doppler} and all the
results are summarized in Table~\ref{lifetimes}. 
\begin{table}[h!]
\begin{center}
\caption[h]{\label{lifetimes} Lifetimes derived in the present work.}
\begin{tabular}{|cc|c|c||cc|c|c|}
\hline
\multicolumn{2}{|c|}{Level}&\multicolumn{2}{c||}{Decay}&\multicolumn{2}{c|}{Level}&\multicolumn{2}{c|}{Decay}\\
E$_i$ &	I$^\pi_i$ &  E$_\gamma$ & $\tau$$^{(a)}$ &	E$_i$  & I$^\pi_i$ &  E$_\gamma$  & $\tau$$^{(a)}$	\\
(keV)&     	 & (keV)       &  (ps)        &     (keV)&             & (keV)       &  (ps) 	\\
\hline
    1744  &4$^-$ 	& 	  612.3  &  0.48(15)&     2667  &5$^{(+)}$    &	  563.8  &  $\le$ 1.4\\
    1751  &8$^-$	&         276.1  &  0.48(20)&     2860  &5$^{(+)}$    &	  757.2  & $\le$ 1.4\\
    1787  &6$^-$	&         432.3  & 0.45(8) &    2864  &7$^{(+)}$    &	  490.2  &$\le$ 0.55\\
    1946  &4$^-$	&         813.6  &  0.47(15)&   2867  &9$^{(+)}$    &	  397.7  &$ \le$ 1.4\\
    2016  &6$^-$	&         661.3  &  0.49(16)&   3155  &7$^{(+)}$	&         780.4  &  0.12(6)  \\
    2465  &10$^-$	&         633.0  & 0.61(16)&    3210  &9$^{(+)}$    &	  740.2  & $\le$ 1.4\\
\hline
\end{tabular}
\end{center}
$^{(a)}$ the number in parenthesis is the error in the last digit.
\end{table}

The spin and parity values of the excited states have been assigned
with the help of the analysis of the $\gamma$-ray angular distributions 
measured 
for the 13 rings of the Euroball IV array. For transitions having too 
weak intensity to be analyzed in that way, their anisotropies have been 
determined using the intensities measured at
two angles relative to the beam axis, 
R$_{ADO}$=I$_{\gamma}$(39.3$^\circ$)/ I$_{\gamma}$(76.6$^\circ$). 
For instance, the angular properties of the 810-, 971- and 
1020-keV transitions, as well as of the 587- and 359-keV
transitions indicate that they are dipole transitions linking states 
with $\Delta I = 1$, such as the 577-keV yrast transition. Then 
the states at 1537, 2103, and 2374~keV have  
odd spin values, while the 2016- and 1787-keV states have even 
spin values.
The $a_2$ angular coefficients of
the 432- and 661-keV transitions (marked with an asterisk in 
Fig.~\ref{schema_part_212Po}) are positive. Since they 
cannot be quadrupole transitions linking states with  
$\Delta I = 2$ because of the other decay 
paths of states located
above them, they are assigned as dipole $\Delta I = 0$ transitions. 
Thus the spin value of the 1787- and 2016-keV states is 6~$\hbar$. These
states cannot have positive parity because their decays towards 
the 4$^+$ yrast state are not observed, whereas they should be 
favored by the energy factors. Then the multipolarities of the 
432- and 661-keV transitions are assigned as E1.
In summary, the spin and parity values of all the
states of $^{212}$Po observed in the present work have been 
determined using similar arguments, both 
from the properties of the populating and decaying 
transitions. As a result, all the transitions given in Table~\ref{lifetimes} 
are assigned to be E1, $\Delta I = 0$.
Their short lifetimes lead to 
{\it very enhanced transitions}, with values of the B(E1) 
reduced transition 
probabilities in the range [2$\times$10$^{-2}$~-~1$\times$10$^{-3}$]~W.u.  
(typical B(E1) values are  $<$ 10$^{-5}$~W.u.).

\begin{figure}[h!]
\begin{center}
\includegraphics[width=7cm]{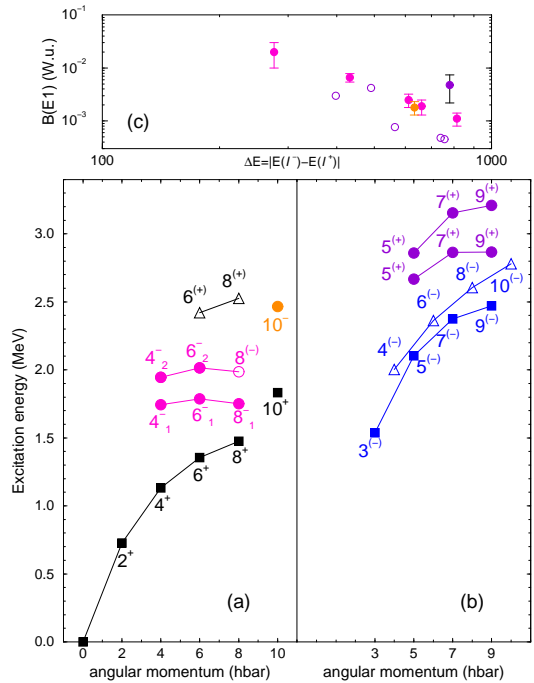}
\caption{(Color online) (a) and (b): Excitation energy of the
$^{212}$Po states as a function of 
angular momentum. Each state drawn 
with a filled circle
decays by an enhanced E1 transition towards
the yrast state having the same I value (drawn with a filled square).
(c): Experimental $B(E1)$ values (in W.u.) versus 
the energy difference (in keV) between $I^{\mp}$ levels (Log-Log plot). 
The empty circles are lowest limits of $B(E1)$, calculated from the
limit values of the lifetimes given in the right part of 
Table~\ref{lifetimes}.}
\label{etats_cluster}
\end{center}
\end{figure}

The bottom part of Fig.~\ref{etats_cluster} displays most of the
$^{212}$Po states observed in this work, grouped as a
function of their underlying structure.
As said in the introduction, low-lying SM configurations account reasonnably well for the
excitation energies of the positive-parity yrast states (drawn with 
filled squares in Fig.~\ref{etats_cluster}(a)). However they fail to
reproduce the large B(E2) transition strengths, while a better description
is achieved using the cluster model~\cite{oh95,bu96}. 
As for the negative-parity
states drawn in Fig.~\ref{etats_cluster}(b), 
they involve the coupling of the low-lying 3$^-$ 
octupole vibration to the excitation of 
the valence nucleons. Such a coupling is well known in the region of 
$^{146}_{~64}$Gd$^{}_{82}$, where the octupole excitation also plays an
important role. It is worth pointing out that the negative-parity 
states shown in Fig.~\ref{etats_cluster}(b) have the same
behavior (relative energies of even- and odd-I states, and de-excitation
modes) as those identified in $^{148}$Gd~\cite{pi90}. 

On the other hand, the two groups of even-I negative-parity states, 
as well as the two groups of 
odd-I positive-parity states (see the filled circles in 
Fig.~\ref{etats_cluster}(a) and Fig.~\ref{etats_cluster}(b), respectively)
cannot be explained by low-lying SM configurations, they are the
fingerprints of the '$\alpha$+$^{208}$Pb' structure, as explained
now.

First of all, strongly enhanced B(E1) values 
are commonly found in nuclei exhibiting an 
{\it electric dipole moment}, such as  
light nuclei described in terms of a bimolecular system rotating
about its center of mass, or heavy nuclei displaying 
octupole deformation~\cite{buna96}. More generally, when a nucleus
clusterizes into fragments with different charge to mass ratios, 
its center of mass does not coincide any more with its center 
of charge, and a sizeable static E1 moment may arise in the intrinsic
frame~\cite{ia85}. For instance the dipole moment of the $\alpha$+$^{208}$Pb
system, computed using Eq.~(1) of Ref.~\cite{ia85}, is as large as 3.7 
$e$fm, that would lead to very enhanced E1 transitions. 
It is worth noting that the E1 rates measured in the present
work (see Fig.~\ref{etats_cluster}(c)) are close to the ones
measured in some heavy nuclei~\cite{ah93}.

Following the ansatz of Ref.~\cite{va92} dealing with the ground state of
$^{212}$Po, we assume that the wavefunction of every excited state 
can be written as the sum of two parts: 
$\Psi^{tot}(I^\pi) =  a~\Psi^{SM}(I^{\pi}) + 
b~\Psi^{cluster}(I^\pi)$. 

Let us discuss the $\alpha$+$^{208}$Pb system. 
Its Hamiltonian has the following form,
$H_{cluster} = H_{\alpha}(r) + H_{dipole}(d)$.
$H_{\alpha}(r)$ is a central potential in which the $\alpha$ 
particle moves (see, e.g.,~\cite{oh95,bu96}), 
its eigenfunctions $\chi_\alpha^\pi (I)$  
have the natural parity, $\pi$ = (-1)$^I$. $H_{dipole}(d)$ is a 
one-dimensional double-well 
potential where the 'collective' coordinate is the dipole moment. 
This double well 
accounts for the fact that in the intrinsic frame, the
$\alpha$-particle can either sit on one side or the other of the 
core. This picture is very similar to the octupole case described
in Ref.~\cite{kr69}, however, here with zero deformation as the core 
is spherical.
In each well we have a ground state 
(no node, $v$~=~0) and one excited state (one node, $v$~=~1). They 
may be viewed as states 
where the $\alpha$ particle vibrates against the Pb core. 
Neither right (R) nor 
left (L) states {\it per se} have good parity. 
The projection on good parity gives the four states, 
$\varphi^{\pm}_{0,1} = \frac{1}{\sqrt{2}}[\varphi^{R}_{0,1} \pm  
\varphi^{L}_{0,1}]$.

Thus, the cluster wavefunctions are given by
\begin{equation}
\Psi^{cluster}_{0,1}(I^\pi) = \chi_\alpha^{\pi ^\prime} (I)\otimes 
\varphi^{\pm}_{0,1}
\end{equation}
with $\pi = \pm \pi ^\prime$,  meaning that, for each $I$ value, 
we get two states with $\pi = +$ and two states with $\pi = -$.

Such a scenario accounts well for the experimental findings. 
The non-natural parity states (4$^-$, 6$^-$,  ..., 5$^+$, 7$^+$, ...) 
are likely pure cluster states ($a \sim 0$). On the 
other hand, the
wavefunctions of the natural-parity states should be strongly 
admixed (cf. the mixing predicted for the ground state of $^{212}$Po 
in the calculation of Ref.~\cite{va92}, $(b/a)^2 \sim$ 0.3) since,  
as mentioned above, the 
low-lying SM configurations give rise to states with even-I and 
positive parity, as well as states with odd-I and negative 
parity (when coupled to octupole vibration).
This situation may explain why the four states of same spin 
but alternating parity do not follow the usual $+, -, +, -$ sequence 
in a 1D double well potential. The two states of positive parity 
are spread by this coupling to SM configurations so that the 
final sequence will be $+, -, -, +$. One has also to invoke the 
mixings in order to understand the downward trend of the B(E1) 
values, as seen in Fig.~\ref{etats_cluster}(c).
This calls for more detailed theoretical studies combining both 
shell and cluster configurations.

In summary we have used the transfer of an $\alpha$ particle induced
by an heavy-ion beam at very
low energy, to populate excited states of $^{212}$Po. This has 
revealed two sets of levels with non-natural  
parity, a first one with even-$I$ values around 2~MeV excitation 
energy and a second one with odd-$I$ values
around 3~MeV. These levels only decay to the 
yrast states having the same $I$ value, by very enhanced E1 transitions 
(B(E1) $\sim 2\times10^{-2}$ -- $1\times10^{-3}$~W.u.). 
They are the fingerprints of the 
'$\alpha$+$^{208}$Pb' structure. 
The oscillatory motion of the $\alpha$-core 
distance around the equilibrium position is observed for the first
time.\\
One may speculate that adding more $\alpha$'s to the $^{208}$Pb core, 
like, e.g. two $\alpha$'s to give $^{216}$Rn, may exhibit similar 
physics. For example the two $\alpha$'s may move coherently as a 
$^8$Be and then the present scenario may repeat itself partially, 
or the two $\alpha$'s move independently and, then, more complex 
structures can be expected.

{\bf Acknowledgments} The Euroball project was a collaboration 
between France, the United Kingdom, Germany, Italy, Denmark and Sweden. 
A.A., P.P. and M.-G.P. are very indebted to their colleagues 
involved the EB-02/17 experiment devoted to the fission fragments, 
in which the present data on $^{212}$Po were recorded. This work was 
partially supported by the collaboration agreement 
Bulgarian Academy
of Sciences-CNRS, under contract No 16946, and by the
contract IDEI-119 of the Romanian Ministry of Education 
and Research.

\end{document}